# Tunable cavity modes and light-pulse propagation in diarylethene-based photo-switchable polymeric microcavities

Francesco Scotognella

Dipartimento di Fisica, Politecnico di Milano, Piazza Leonardo da Vinci 32, 20133 Milano, Italy
e-mail address: francesco.scotognella@polimi.it

**Abstract**
Light responsive devices employing molecular photo-switches are interesting for displays and dynamic light filtering. In this work, polymeric microcavities embedding a layer of photochromic compound have been studied by means of the transfer matrix method. Different polymers, such as poly vinyl carbazole, poly styrene and cellulose acetate, have been used. A microcavity that includes random one-dimensional photonic structures sandwiching the photochromic layer has also been studied. Propagation of light pulses through the microcavities has been analysed.

**Keywords:** tunable microcavity; photochromic molecule; light-pulse propagation; transfer matrix method

**Introduction**

Photochromic materials are of great interest because of their employment in light-responsive devices, as for example rewritable displays, dynamic optical filters and optical memories [1–4]. In photochromism, an electromagnetic radiation-induced reversible transformation, from a thermodynamically more stable isomer A into a metastable isomer B, occurs [5]. Recently, neuronal photo-stimulation has been demonstrated with an azobenzene-based photo-switch [6,7].
The modulation of molecular photo-switches can be combined to the optical properties of photonic crystals, in order to achieve photo-modulation of the photonic band gap of the photonic crystal [8,9]. One-dimensional photonic crystals are very versatile systems that can be fabricated with several methods, such as spin coating [10–13], sputtering [14] and pulsed laser deposition [15,16], and that can be employed for sensors [17,18] and lasers [19,20,11]. A one-dimensional microcavity can be obtained by placing a defect layer between two one-dimensional photonic crystals [21,22].
In this work, a defect layer of the molecular photo-switch diarylethene1 [23,24] has been embedded between two polymeric one-dimensional photonic crystals, made alternating either poly vinyl carbazole and poly styrene or poly vinyl carbazole and cellulose acetate. A microcavity, in which the defect includes random photonic crystals that embed the diarylethene1 layer, is also studied. The optical properties of the microcavities has been simulated via the transfer matrix method, considering the refractive index dispersions of all the materials. The photo-stimulation of diarylethene1 leads to a modulation of the cavity modes. The transmission of a light pulse through the microcavities has been studied, highlighting a temporal delay in correspondence of the cavity modes.

**Methods**

The transfer matrix method is a powerful tool to describe the optical properties of the one-dimensional photonic structures [25–29]. The studied system is glass/multilayer/air with light that

impinges the sample orthogonally with respect to the sample surface. Within this framework the parameter

$$\phi_k(\lambda) = \frac{2\pi}{\lambda} n_k(\lambda) d_k \qquad (1)$$

is used the characteristic matrix for the *k*th layer [25]

$$M_k = \begin{bmatrix} \cos\phi_k(\lambda) & -\frac{i}{n_k(\lambda)} \sin\phi_k(\lambda) \\ -i n_k(\lambda) \sin\phi_k(\lambda) & \cos\phi_k(\lambda) \end{bmatrix} \qquad (2)$$

with $n_k(\lambda)$ and $d_k$ the wavelength-dependent refractive index and the thickness (in nm), respectively, of the *k*th layer. The matrix for the multilayer is given by

$$M = \begin{bmatrix} M_{11} & M_{12} \\ M_{21} & M_{22} \end{bmatrix} = \prod_{i=1}^{N} M_k \qquad (3)$$

With *N* the number of layers. From the matrix *M* it is possible to determine the transmission coefficient *t*:

$$t = \frac{2n_s}{(M_{11} + M_{12} n_0) n_s + (M_{21} + M_{22} n_0)} \qquad (4)$$

And the transmission *T*:

$$T = \frac{n_0}{n_s} |t|^2 \qquad (5)$$

$n_s$ and $n_0$ are the refractive indexes of the substrate (glass, $n_s = 1.46$) and air, respectively.

The temporal response of the systems has been studied by employing the inverse fast Fourier transform of the product of the incoming light pulse and the transmission spectrum of the multilayers. The transmitted time resolved signal $T(t)$ can be written as

$$T(t) = \mathcal{F}^{-1}[G(\omega) T(\omega)] = \int_{-\infty}^{\infty} G(\omega) T(\omega) e^{i\omega t} d\omega \qquad (6)$$

In which $G(\omega)$ is the Gaussian spectral profile of the incoming light pulse and $T(\omega)$ is the transmission spectrum of the multilayers [30].

The multilayers are made with different materials, such as diarylethene1 [24], poly vinyl carbazole (PVK), poly styrene (PS), cellulose acetate (CA). The complex refractive index of diarylethene1, in its open and close form, is reported in Ref. [31]. The Sellmeier equation for the refractive index of PVK is given by [13,32]:

$$n_{PVK}^2(\lambda) - 1 = \frac{0.09788\lambda^2}{\lambda^2 - 0.3257^2} + \frac{0.6901\lambda^2}{\lambda^2 - 0.1419^2} + \frac{0.8513\lambda^2}{\lambda^2 - 1.1417^2} \qquad (7)$$

The Sellmeier equation for the refractive index of PS is given by [33,34]:

$$n_{PS}^2(\lambda) - 1 = \frac{1.4435\lambda^2}{\lambda^2 - 0.020216^2} \qquad (8)$$

Finally, the Sellmeier equation for the refractive index of CA is [13,32]:

$$n_{CA}^2(\lambda) - 1 = \frac{0.6481\lambda^2}{\lambda^2 - 0.0365^2} + \frac{0.5224\lambda^2}{\lambda^2 - 0.1367^2} + \frac{2.483\lambda^2}{\lambda^2 - 13.54^2} \qquad (9)$$

In the Sellemeier equations $\lambda$ is in micrometers.

**Results and Discussion**

In this work different microcavities that embed a diarylethene1 layer between two polymeric one-dimensional photonic crystals are studied. The one-dimensional photonic crystals are made by alternating either layers of PVK and PS or layers of PVK and CA. Moreover, it has been studied a microcavity in which the defect is made by the diarylethene layer sandwiched by two random photonic crystal of 20 layers of PVK and PS; the defect is embedded between two one-dimensional photonic crystals of PVK and CA.

The first microcavity is made with a diarylethene1 defect layer between two photonic crystals of 40 bilayers of PVK and PS. Thus, the studied multilayer is

glass/(PVK/PS)$_{40}$/diarylethene1/(PS/PVK)$_{40}$/air. The thickness of the PVK layers is 75.2 nm, the thickness of the PS layers is 79.9 nm, and the thickness of diarylethene1 defect layer is 150.4 nm. In Figure 1 the transmission spectra of the microcavity (PVK/PS)$_{40}$/diarylethene1/(PS/PVK)$_{40}$ with diarylethene1 in the open form (solid black curve) and in the closed form (dashed red curve) are shown.

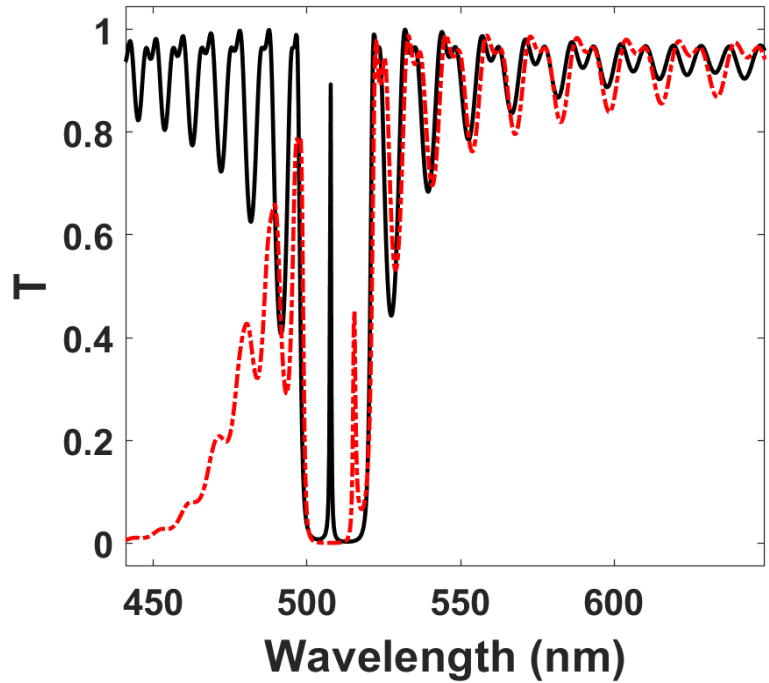

**Figure 1.** Transmission spectra of the microcavity (PVK/PS)$_{40}$/diarylethene1/(PS/PVK)$_{40}$ with diarylethene1 in the open form (solid black curve) and in the closed form (dashed red curve).

With the photo-stimulation the complex refractive index of diarylethene1 is changing, inducing a shift of the cavity mode in the microcavity. In Figure 2 (above) the transmission spectra of the microcavity (PVK/PS)$_{40}$/diarylethene1/(PS/PVK)$_{40}$, with diarylethene1 in the open form (solid black curve) and in the closed form (dashed red curve) are shown together with the spectral shapes of three light pulses centred at 545, 580 and 589 THz. The width of the (Gaussian) peaks is 1 THz. With equation 6 the time-resolved transmitted signals through the microcavity (Figure 2, below). The duration of the pulse out of the cavity (545 THz) is 1 picosecond, while the pulses at 580 THz (for Diarylethene1 in the open form) and at 589 THz (for Diarylethene1 in the closed form) are significantly delayed.

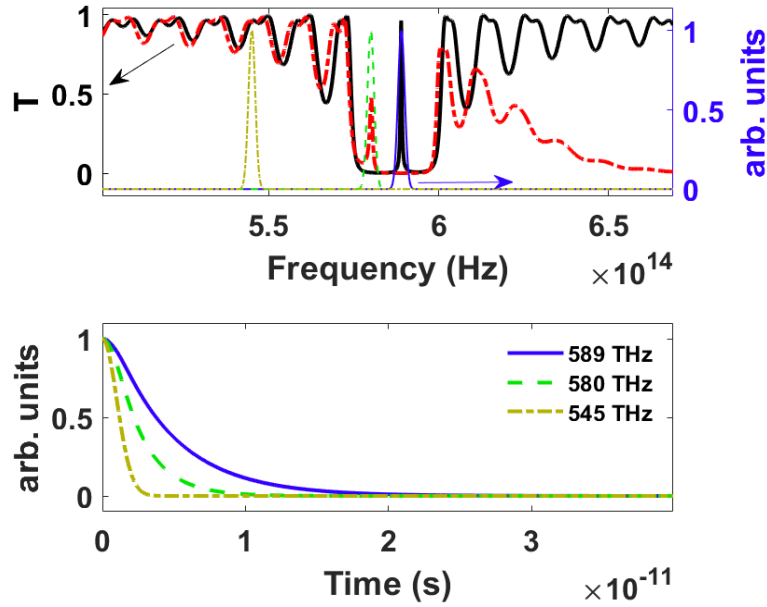

**Figure 2.** (above) Transmission spectra of the microcavity (PVK/PS)$_{40}$/diarylethene1/(PS/PVK)$_{40}$, with diarylethene1 in the open form (solid black curve) and in the closed form (dashed red curve), and spectral shapes of three light pulses centred at 545, 580 and 589 THz. (below) Time-resolved transmitted signals through the microcavity.

The second studied system is glass/(PVK/CA)$_{25}$/diarylethene1/(CA/PVK)$_{25}$/air. The thickness of the PVK layers is 84 nm, the thickness of the CA layers is 95.2 nm, and the thickness of diarylethene1 defect layer is 172.8 nm. In Figure 3 the transmission spectra of the microcavity (PVK/CA)$_{25}$/diarylethene1/(PS/PVK)$_{25}$ with diarylethene1 in the open form (solid black curve) and in the closed form (dashed red curve) are shown. With respect to the microcavity with PVK and PS the chosen number of bilayers of the two photonic crystals is smaller. Nevertheless, the transmission in the photonic band region is close to zero in both microstructures. This is due to refractive index contrast, which is higher in the case of PVK and CA (at about 600 nm, $\Delta n_{PVK-CA} = n_{PVK} - n_{CA} \cong 0.183$, $\Delta n_{PVK-PS} = n_{PVK} - n_{PS} \cong 0.074$, respectively).

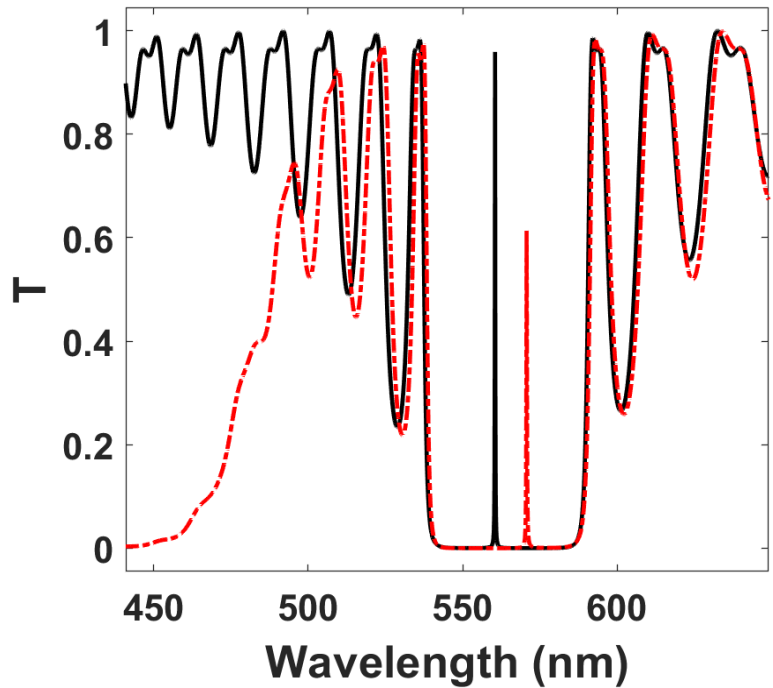

**Figure 3.** Transmission spectra of the microcavity (PVK/CA)$_{25}$/diarylethene1/(PS/PVK)$_{25}$ with diarylethene1 in the open form (solid black curve) and in the closed form (dashed red curve).

In Figure 3 (above) the transmission of the microcavity is reported together with the spectral shapes of three light pulses at 470, 524 and 533 THz. As in the microcavity with PVK and PS, the light pulses in correspondence of the cavity modes, for diarylethene1 in the open and closed forms, are delayed (Figure 3, below).

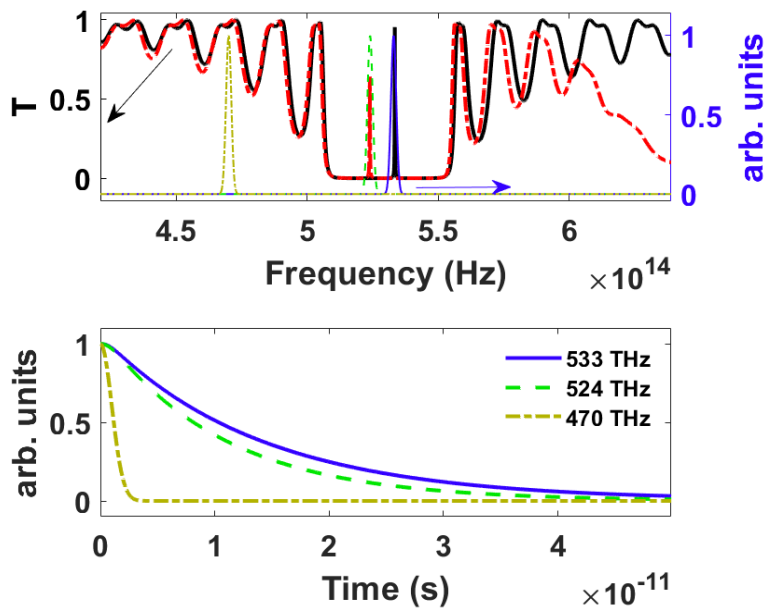

**Figure 4.** (above) Transmission spectra of the microcavity (PVK/CA)$_{25}$/diarylethene1/(CA/PVK)$_{25}$, with diarylethene1 in the open form (solid black curve) and in the closed form (dashed red curve), and spectral shapes of three light pulses centred at 470, 524 and 533 THz. (below) Time-resolved transmitted signals through the microcavity.

The time delay of the two pulses can be tuned by varying the number of the layers in the microcavity. In Figure 5 the delay time of the transmitted light pulses through the (PVK/CA)$_{NL}$/diarylethene1/(CA/PVK)$_{NL}$ microcavity, corresponding to 50% of the intensity of the initial signal, as a function of the number of layers *NL* is shown.

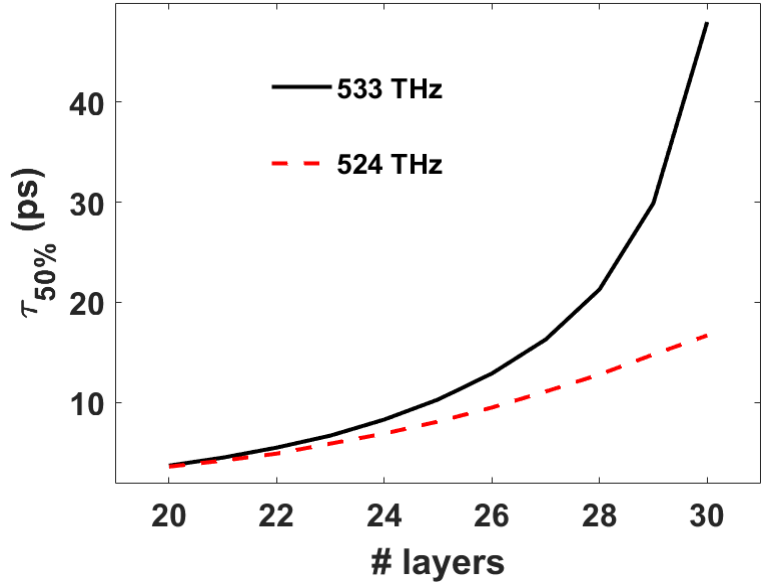

**Figure 5.** Delay time of the transmitted light pulses through the (PVK/CA)$_{NL}$/diarylethene1/(CA/PVK)$_{NL}$ microcavity, corresponding to 50% of the intensity of the initial signal, as a function of the number of layers *NL* (black solid curve is related to diarylethene1 in the open form, red dashed curve is related to diarylethene1 in the closed form).

In the third microcavity the defect layer includes two PVK/PS random photonic sequences, of 20 layers, sandwiching the diarylethene1 layer. The defect is embedded between two PVK/CA one-dimensional photonic crystals. Thus, the structure of the microcavity is (PVK/CA)$_{25}$/PVK/PS/PS/PS/PVK/PS/PS/PVK/PS/PVK/PS/PVK/PS/PVK/PS/PVK/PVK/PVK/PVK/PS/diarylethene1/PVK/PS/PS/PS/PS/PVK/PS/PS/PS/PS/PS/PVK/PS/PVK/PVK/PVK/PS/PS/PVK/PVK/(CA/PVK)$_{25}$. The thickness of the PVK layers is 84 nm, the thickness of the PS layers is 89.6 nm, the thickness of the CA layers is 95.2 nm, and the thickness of diarylethene1 defect layer is 172.8 nm.

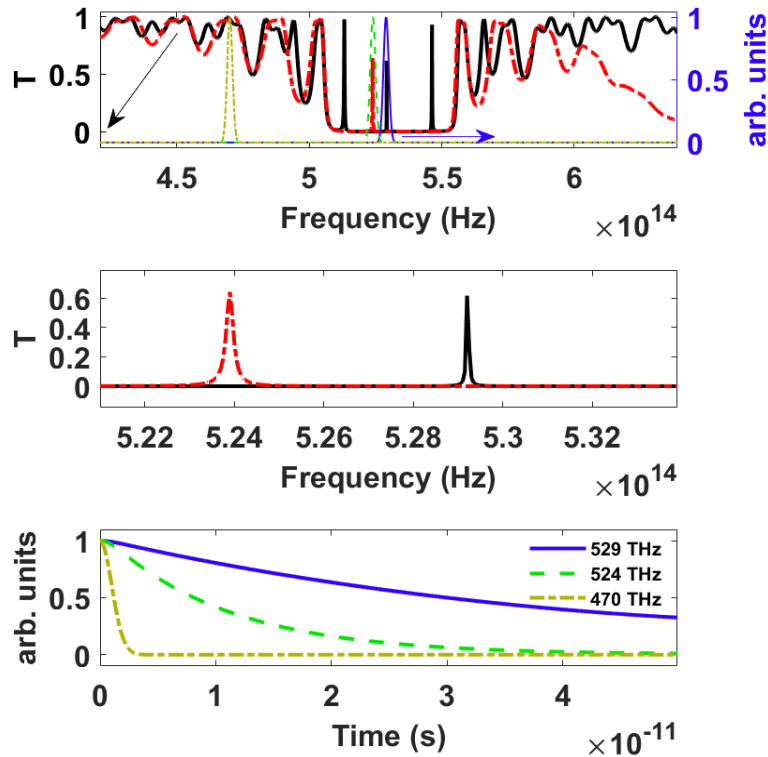

**Figure 6.** (above) Transmission spectra of the microcavity (PVK/CA)$_{NL}$/(randomPVK/PS)/diarylethene1/(randomPVK/PS)/(CA/PVK)$_{NL}$, with diarylethene1 in the open form (solid black curve) and in the closed form (dashed red curve), and spectral shapes of three light pulses centred at 470, 524 and 529 THz. (middle) Magnification of the transmission spectrum in the region of the central cavity modes. (below) Time-resolved transmitted signals through the microcavity.

The delay time for the transmitted signals at 529 THz, for the microcavity with diarylethene1 in the open form, and at 524 THz, for the microcavity with diarylethene1 in the closed form, is significantly different. This is due to the fact that the disorder in the microcavity leads to a broader cavity mode in the case of diarylethene1 in the closed form (dashed red curve in Figure 6, middle).

With the PVK/PS microcavity a shift, from the cavity mode for diarylethene1 in the open form to the one with the dye in the closed form, related to $\Delta\lambda/\lambda = 0.015$ has been achieved, while with the PVK/CA microcavity a shift related to $\Delta\lambda/\lambda = 0.018$ has been achieved. With the PVK/CA random sequence microcavity, the shift corresponds to a $\Delta\lambda/\lambda = 0.010$. In literature, experiments on (PVK/CA)$_{15}$/PVA/PMA4/PVA/(CA/PVK)$_{15}$ report a $\Delta\lambda/\lambda = 0.011$ [35,36].

**Conclusion**

In this study different microcavities in which a diarylethene1 layer is sandwiched between two polymeric one-dimensional photonic crystals have been simulated by means of the transfer matrix method. The one-dimensional photonic crystals have been designed by alternating either layers of PVK and PS or layers of PVK and CA. Furthermore, it has been analysed a microcavity in which the defect layer of diarylethene1 is sandwiched between two random photonic crystals of 20 layers of PVK and PS. In such microcavity the defect (of 41 layers in total) is embedded between two one-dimensional photonic crystals of PVK and CA. Light pulse propagation through these microcavities have been also analysed, highlighting a modulation of the light pulse delay out of the cavity band gap and in the cavity modes with diarylethene1 in the open and closed forms. Such findings can be interesting for the fabrication of dynamic light filters via tunable structural colour and optical memories.

**Acknowledgement**

This work was supported by Grant No. CFPMN1–008 from the Department of National Defence Discovery supplement and IDEAS.